\documentclass[prl,twocolumn,floatfix,superscriptaddress]{revtex4-1}
\usepackage{dcolumn,amsmath}
\usepackage{graphicx}
\usepackage{bm}
\usepackage{hyperref}

\newcommand{\ecm}{\ensuremath{e {\cdotp} {\rm cm}}}
\newcommand{\eEDM}{{\em e}EDM}

\newcommand{\B}{\mathcal{B}} %magnetic field
\newcommand{\E}{\mathcal{E}} %electric field
\newcommand{\de}{d_\mathrm{e}}

\newcommand{\Dp}{D_{\|}}
\newcommand{\Brot}{B_\mathrm{rot}}

\begin{document}
\title{Zeeman interaction in $^3\Delta_1$ state of HfF$^+$ to search for the electron electric-dipole-moment}
%\title{Suppression of systematics in the eEDM experiment }

\author{A.N.\ Petrov}\email{alexsandernp@gmail.com}
\author{L.V.\ Skripnikov}\email{leonidos239@gmail.com}
\author{A.V.\ Titov}
\homepage{http://www.qchem.pnpi.spb.ru}
\affiliation{National Research Centre ``Kurchatov Institute'' B.P. Konstantinov Petersburg Nuclear Physics Institute, Gatchina, Leningrad District 188300, Russia}
\affiliation{Saint Petersburg State University, 7/9 Universitetskaya nab., St. Petersburg, 199034 Russia}
\date{\today}

\begin{abstract}
We report the theoretical investigation of the suppression of magnetic systematic effects in HfF$^+$ cation for the experiment to search for the electron electric dipole moment. 
%ap
The g-factors for $J = 1$, $F=3/2$, $|M_F|=3/2$ hyperfine 
levels of the $^3\Delta_1$ state are calculated as functions of the external electric field. The lowest
value for the difference between the g-factors of $\Omega$-doublet levels, 
$\Delta g = 3 \times 10^{−6}$, is attained at the electric field 7 V/cm.
The body-fixed g-factor, $G_{\parallel}$, was obtained both within the electronic structure calculations and with our fit of the experimental data from [H. Loh, K. C. Cossel, M. C. Grau, K.-K. Ni, E. R. Meyer, J. L. Bohn, J. Ye, and E. A. Cornell, Science {\bf 342}, 1220 (2013)]. 
%end ap
For the electronic structure calculations we used a combined scheme to perform correlation calculations of HfF$^+$ which includes both the direct 4-component all-electron and generalized relativistic effective core potential approaches. The electron correlation effects were treated using the coupled cluster methods.
%ap 
The calculated value $G_{\parallel}=0.0115$ agrees very well with the $G_{\parallel}=0.0118$
obtained in the our fitting procedure.
%end ap
%ap
%Final values of the molecule frame dipole moment (with the origin in the center of mass) and g-factor in the working $^3\Delta_1$ state are $-3.78$ D and $0.0115$, respectively. 
The calculated value $D_{\parallel}=-1.53$ a.u. of the molecule frame dipole moment (with the origin in the center of mass) is in agreement with the experimental value $D_{\parallel}=-1.54(1)$ a.u.
[H. Loh, Ph.D. thesis, Massachusetts Institute of Technology (2006)].
%end ap

\end{abstract}

\maketitle

\section{Introduction}
Search for the electron electric dipole moment (\eEDM), $d_e$, is important test of the standard model and its extensions \cite{Commins:98,Chupp:15}. The best current limit on the electron EDM, $|\de|<9\times 10^{-29}$ \ecm\, was set with a molecular beam of the thorium monoxide (ThO) molecules by ACME collaboration \cite{ACME:14a}
%ap1
using the theoretical data from Refs. \cite{Skripnikov:13c,Skripnikov:15a,Skripnikov:16b}.
%end ap1
 A number of other systems are considered to search for the \eEDM\ and other manifistations of effects of time-reversal (T) and spatial parity (P) symmetries violation of the fundamental interactions:  ThO~\cite{FDK14,Skripnikov:13c,Skripnikov:14a,Skripnikov:15a,Skripnikov:16b}, TaN\cite{FDK14,Skripnikov:15c}, ThF$^+$~\cite{Cornell:13,Skripnikov:15b}, 
PbF \cite{Shafer-Ray:08E, Skripnikov:14c, Petrov:13, Skripnikov:15d}, WC \cite{Lee:13a, Meyer:09a}, RaO \cite{Flambaum:08,Kudashov:13}, RaF \cite{Isaev:12,Kudashov:14}, PtH$^+$ \cite{Meyer:06a, Skripnikov:09}, etc.), TlF~\cite{Skripnikov:09a,Petrov:02,Hinds:80a,Laerdahl:97} molecules and cations.  

E. Cornell's group has suggested to use the trapped molecular ions for the \eEDM\ search \cite{Meyer:06a,Meyer:08}. One of the most promising systems for the search is the HfF$^+$ cation 
\cite{Cossel:12, Cornell:13, Petrov:07a, Petrov:09b, Fleig:13, Meyer:06a, Le:13}
which is also of interest for other fundamental experiments~\cite{Skripnikov:08a,Skripnikov:17b}.
 It has the long-lived metastable $^3\Delta_1$ electronic state with lifetime $\approx 2 s$ \cite{Petrov:09b,Cossel:12} which means a very large coherence time is achievable in the experiment.
%%
%ap: moved down
%
%end ap
%.....
The other main feature of the $^3\Delta_1$ state is that it has a very small g-factor (zero in nonrelativistic limit in approximation with free-electron g-factor,
%ap
g$_{S}$,
%end ap
 equal to -2.0) which leads to the suppression of the magnetic systematic effects. 
%ap
It was shown that further suppression of systematics is possible due to existence of the $\Omega$-doublet structure of molecules in the the $^3\Delta_1$ electronic state \cite{DeMille2001, Petrov:14, Vutha2009, Petrov:15}.
%end ap
%
For preparation and 
%ap1
%conduction
implementation
%end ap1
 of the \eEDM\ experiment one should investigate the dependence of upper and lower $\Omega$-doublet states
%ap
g-factors
%end ap
 on the strength of the laboratory electric field.
%...
And this is the goal of the present paper.

\section{Theory}
We define the $g$-factors such that Zeeman shift is equal to
\begin{equation} 
%   E_{\rm Zeeman}(\rm{\E}) = -g(\rm{\E})\mu_B \B M.
   E_{\rm Zeeman} = -g \mu_B \B M_F,
 \label{Zeem}
\end{equation}
where $\mu_{B}$ is the Bohr magneton, $M_F$ is the projection of the total angular momentum on the lab $z$ axis, ${\bf B} = \B\hat{z}$ is the external magnetic field.
This definition matches the ones in the papers \cite{Petrov:14,Cornell:13}.
%ap1
Using the angular momentum algebra \cite{LL77}, one can calculate that
%end ap1
 in the adiabatic approximation and in the limit of zero hyperfine interaction
 $g$-factors of hyperfine sublevels of the $^3\Delta_1$ state
of  HfF$^+$ are determined by

\begin{eqnarray}
 g = -G_{\parallel}\frac{F(F+1)+J(J+1)-3/4}{2F(F+1)J(J+1)} + \\
g_F\frac{\mu_{N}}{\mu_{B}}\frac{F(F+1)-J(J+1)+3/4}{2F(F+1)}
\label{geq}
\end{eqnarray} 
Here $g_F=5.25773$ is $^{19}$F nucleus $g-$factor,  $\mu_{N}$ is the nuclear magneton. The first term in the right hand side of Eq. (\ref{geq})
is the electronic contribution \cite{Petrov:11} and the second term is contribution from the magnetic moment of $^{19}$F nucleus.

Eq. (\ref{geq}) does not take into account the hyperfine interaction between different rotational levels and nonadiabatic interaction with other electronic states.
To take these effects into account, following
Refs. \cite{Petrov:11,Petrov:14}, the $g$-factors are obtained by numerical diagonalization of the molecular Hamiltonian (${\rm \bf \hat{H}}_{\rm mol}$) in external electric ${\bf E} = \E\hat{z}$ and magnetic ${\bf B} = \B\hat{z}$ fields 
%ap: see below
over the basis set of the electronic-rotational wavefunctions
\begin{equation}
 \Psi_{\Omega}\theta^{J}_{M,\Omega}(\alpha,\beta)U^{\rm F}_{M_I}.
\label{basis}
\end{equation}
Here $\Psi_{\Omega}$ is the electronic wavefunction, $\theta^{J}_{M,\Omega}(\alpha,\beta)=\sqrt{(2J+1)/{4\pi}}D^{J}_{M,\Omega}(\alpha,\beta,\gamma=0)$ is the rotational wavefunction, $\alpha,\beta,\gamma$ are Euler angles, $U^{F}_{M_I}$ is the F nuclear spin wavefunctions and $M$ $(\Omega)$ is the projection of the molecule angular momentum on the lab $\hat{z}$ (internuclear $\hat{n}$) axis, $M_I=\pm1/2$ is the projection of the nuclear angular 
momentum on the same axis. Note that $M_F=M_I+M$.

%ap
We represent the molecular Hamiltonian for $^{180}$Hf$^{19}$F$^+$ as
\begin{equation}
{\rm \bf\hat{H}}_{\rm mol} = {\rm \bf \hat{H}}_{\rm el} + {\rm \bf \hat{H}}_{\rm rot} + {\rm \bf\hat{H}}_{\rm hfs} + {\rm \bf\hat{H}}_{ext} .
\end{equation} 
Here ${\rm \bf \hat{H}}_{\rm el}$ is the electronic Hamiltonian,
\begin{equation}
 {\rm \bf\hat{H}}_{\rm rot} = \Brot { {\bf J}}^2 -2\Brot({ {\bf J}}\cdot{ \vec{\bf J}}^e)
\end{equation}
is the rotational Hamiltonian,
\begin{equation}
 {\rm \bf\hat{H}}_{\rm hfs} = g_{\rm F} \bm{I} \cdot \sum_i\left(\frac{\bm{\alpha}_i\times \bm{r}_i}{r_i^3}\right)
\end{equation}
is the hyperfine interaction between electrons and flourine nuclei,
\begin{equation}
 {\rm \bf\hat{H}}_{\rm ext} = \mu_{\rm B}({ {\bf L}}^e-g_{S}{ {\bf S}}^e)\cdot{\bf B} -g_{\rm F}\frac{\mu_{N}}{\mu_{B}}\bm{I}\cdot{\bf B}   -{ {\bf D}} \cdot {\bf E}
\end{equation}
describes the interaction of the molecule with external magnetic and electric fields,
$\Brot=0.2989$ \cite{Cossel:12} is the rotational constant, $ g_{S} = -2.0023$ is a free$-$electron $g$-factor,
{\bf D} is the dipole moment operator.
%end ap

For the current study we have considered the following low-lying electronic basis states: $^3\Delta_1$,  $^3\Delta_2$,  $^3\Pi_{0^+}$ and $^3\Pi_{0^-}$.
%ap
 ${\rm \bf \hat{H}}_{\rm el}$ is diagonal on the basis set (\ref{basis}). Its eigenvalues are
 transition energies of these states. They were calculated and measured in Ref.~\cite{Cossel:12}:
\begin{align}
\label{Molbasis}
\nonumber
^3\Delta_1 & : T_e=976.930~{\rm cm}^{-1}\ ,\\
\nonumber
 ^3\Delta_2 & : T_e=2149.432~{\rm cm}^{-1}\ ,\\
 \nonumber
^3\Pi_{0^-} & : T_e=10212.623~{\rm cm}^{-1}\ ,\\
 %\nonumber
 ^3\Pi_{0^+} & : T_e=10401.723~{\rm cm}^{-1}\ .
%\end{eqnarray}
\end{align}
%end ap
%
Other terms of molecular Hamiltonian ${\rm \bf \hat{H}}_{\rm mol}$ are determined by parameters given by Eqs. (\ref{Gperp1})--- (\ref{Gpar}) below.
%To construct the matrix of the molecular Hamiltonian 
%
We have performed electronic calculations for the following matrix elements of the basis electronic states:

\begin{eqnarray}
 \label{Gperp1}
   G_{\perp}^{(1)} &=& \langle  ^3\Delta_1  |\hat{L}^e_- -  g_{S}\hat{S}^e_-|^3\Delta_2  \rangle = -2.617, \\
 \label{dip1}
   D_{\perp}^{(1)} &=& \langle  ^3\Delta_1  |\hat{D}_-  |^3\Delta_2  \rangle =  0.034~ {\rm a.u.}, \\
\label{Delt1}
    \Delta^{(1)}  &=&	2\Brot\langle ^3\Delta_1 |J^e_- | ^3\Delta_2 \rangle = -.7370874~ {\rm cm}^{-1},\\
 \label{dip}
   \Dp &=& \langle ^3\Delta_1 |\hat{D}_{\hat{n}} |^3\Delta_1 \rangle=-1.53~ {\rm a.u.},
\end{eqnarray}
%ap
\begin{eqnarray}
 \label{dip2}
\label{transdip2a}
D_{\perp}^{(2a)} = \langle ^3\Delta_1  | \hat{D}_+  | ^3\Pi_{0^+} \rangle =  0.457~ {\rm a.u.}, \\
\label{transdip2b}
D_{\perp}^{(2b)} = \langle ^3\Delta_1  | \hat{D}_+  | ^3\Pi_{0^-} \rangle =  0.447~ {\rm a.u.}.
\end{eqnarray}
Matrix element (\ref{transdip2a}) is in a good agreement with the value $D_{\perp}^{(2a)} = 0.467$ a.u. calculated in Ref. \cite{Petrov:09b}.
Calculated permanent dipole moment $\Dp$ is in a good agreement with the experimental value  $\Dp=-1.54(1)$ a.u. \cite{Loh2006Thesis}.
%end ap
%where $ g_{S} = -2.0023$ is a free$-$electron $g$-factor.
Matrix elements
\begin{eqnarray}
 \label{Gperp2a}
   G_{\perp}^{(2a)} &=& \langle ^3\Delta_1  |\hat{L}^e_+ -  g_{S}\hat{S}^e_+| ^3\Pi_{0^+} \rangle  = 1.3456, \\
 \label{Gperp2b}
   G_{\perp}^{(2b)} &=& \langle ^3\Delta_1  |\hat{L}^e_+ -  g_{S}\hat{S}^e_+| ^3\Pi_{0^-} \rangle = 1.5524, \\
 \label{Delt2a}
    \Delta^{(2a)}  &=&	2\Brot\langle ^3\Delta_1 |J^e_+ | ^3\Pi_{0^+} \rangle = .8044~  {\rm cm}^{-1}, \\
 \label{Delt2b}
    \Delta^{(2b)}  &=&	2\Brot\langle ^3\Delta_1 |J^e_+ | ^3\Pi_{0^-}  \rangle = .9280~  {\rm cm}^{-1}
\end{eqnarray}
were chosen in such a way to reproduce the experimental value $0.369\cdot J(J+1)$ MHz for $\Omega$ doubling of $^3\Delta_1$.
Matrix element
\begin{equation}
 \label{All}
   A_{\parallel}= g_{\rm F}
   \langle
   \Psi_{^3\Delta_1}|\sum_i\left(\frac{\bm{\alpha}_i\times \bm{r}_i}{r_i^3}\right)_\zeta|\Psi_{^3\Delta_1}
   \rangle= -58.1~{\rm MHz}
%\end{eqnarray}
\end{equation}
were taken from Ref. \cite{Petrov:09b}.
%ap
Hyperfine structure only of the $^3\Delta_1$ state was taken into account.
%end ap
%ls
%And finally matrix element is obtained both
%The value of the matrix element
G$_{\parallel}$ is given by the following formula:
\begin{eqnarray}
 \label{Gpar}
       G_{\parallel} &=&\frac{1}{\Omega} \langle ^3\Delta_1 |\hat{L}^e_{\hat{n}} - g_{S} \hat{S}^e_{\hat{n}} |^3\Delta_1 \rangle.
\end{eqnarray}
%Otained value $G_{\parallel}=.011768$ is very close to that obtained form eq. (\ref{geq})
%\begin{eqnarray}
% \label{Gpar}
%       G_{\parallel} &=&\frac{1}{\Omega} \langle ^3\Delta_1 |\hat{L}^e_{\hat{n}} - g_{S} \hat{S}^e_{\hat{n}} |^3\Delta_1 \rangle, \\
% \label{Gperp1}
%   G_{\perp}^{(1)} &=& \langle  ^3\Delta_2  |\hat{L}^e_+ -  g_{S}\hat{S}^e_+|^3\Delta_1  \rangle, \\
% \label{Gperp2}
%   G_{\perp}^{(2)} &=& \langle ^3\Delta_1  |\hat{L}^e_+ -  g_{S}\hat{S}^e_+| ^3\Pi_{0^\pm} \rangle, \\
% \label{Delt1}
%    \Delta^{(1)}  &=&	2\Brot\langle ^3\Delta_2 |J^e_+ | ^3\Delta_1 \rangle~, \\
% \label{Delt2}
%    \Delta^{(2)}  &=&	2\Brot\langle ^3\Delta_1 |J^e_+ | ^3\Pi_{0^\pm}  \rangle~, \\
% \label{dip}
%   \Dp &=& \langle ^3\Delta_1 |\hat{D}_{\hat{n}} |^3\Delta_1 \rangle,\\
% \label{dip1}
%   D_{\perp}^{(1)} &=& \langle  ^3\Delta_2  |\hat{D}_+  |^3\Delta_1  \rangle, \\
% \label{dip2}
%   D_{\perp}^{(2)} &=& \langle ^3\Delta_1  | \hat{D}_+  | ^3\Pi_{0^\pm} \rangle,
%\end{eqnarray}
%where $ g_{S} = -2.0023$ is a free$-$electron $g$-factor. 

To perform electronic structure calculations of the diagonal matrix elements (\ref{dip}) and (\ref{Gpar})
 we have used the combined computational scheme similar to that used in Refs.~\cite{Skripnikov:16b,Skripnikov:17a,Skripnikov:17b}  which includes electronic structure treatment within the generalized relativistic effective core (GRECP) potential approach \cite{Mosyagin:10a,Mosyagin:16} and the direct relativistic 4-component Dirac-Coulomb(-Gaunt) approach. This scheme includes the following stages:
(i) 2-component 52-electron relativistic correlation calculation using the coupled cluster with single, double and noniterative triple cluster amplitudes, CCSD(T), method.
For this we have used the semilocal version of the 44-electron GRECP operator~\cite{Mosyagin:10a,Mosyagin:16}.
The 28 inner core $1s..3d$ electrons of Hf have been excluded from the correlation treatment by the GRECP operator and all other (outer core and valence) electrons were included in the correlation calculation.
(ii) To treat the correlation contribution from the inner core electrons we have performed direct 4-component calculations at the level of the coupled cluster with single amplitudes (CCS) method as the difference in the calculated properties within the 80-electron (i.e. all-electron) CCS versus the 52-electron CCS. 
(iii) Calculation of vibration correction for $G_{\parallel}$..
(iv) Calculation of the correction on high order correlation effects.

For the stage (i) we have generated the 
%ls1
%partly contracted basis set for Hf that includes 15 $s-$, 10 $p-$, 8 $d-$, 7 $f-$, 4 $g-$, 2 $h-$ and 1 $i-$ type Gaussians (only $g-$, $h-$ and $i-$ type basis functions were contracted using the code from Ref.~\cite{Skripnikov:13a}).
uncontracted basis set for Hf that includes 25 $s-$, 25 $p-$, 21 $d-$, 14 $f-$, 10 $g-$, 5 $h-$ and 5 $i-$ type Gaussians
%end ls1
For fluorine the 
%ls1
%(13,7,4,3)/[6,5,4,3]
(13,7,4,3,2)/[6,5,4,3,2]
%end ls1
 aug-ccpVQZ basis set \cite{Dunning:89,Kendall:92} 
was used.
Note that the reduction of the basis set on Hf to 15 $s-$, 10 $p-$, 8 $d-$, 7 $f-$, 4 $g-$, 2 $h-$ and 1 $i-$ type Gaussians (only $g-$, $h-$ and $i-$ type basis functions were contracted using the code from Ref.~\cite{Skripnikov:13a}) 
%ls2
%lead to only 
leads only to 
%end ls2
slight changes in the calculated values.

 For the stage (ii) the CVDZ \cite{Dyall:07,Dyall:12} basis set for Hf and the ccpVDZ \cite{Dunning:89,Kendall:92} basis set for F were used.
In the stage (iv) the high order correlation effects were considered as a difference in the values of considered properties calculated within the coupled cluster with single, double, triple and noniterative quadruple amplitudes and the CCSD(T) method. In the calculations 20 valence and outer core electrons of HfF$^+$ were correlated. 
%end ls1

%As expected inner-core correlation negligibly contribute to the 

To calculate off-diagonal matrix elements 
%(\ref{Gperp1}), (\ref{Gperp2}), (\ref{Delt1}), (\ref{Delt2}), (\ref{dip1}) and (\ref{dip2}) 
%(\ref{Gperp1}), (\ref{Delt1}) and (\ref{dip1})
(\ref{Gperp1}), (\ref{dip1}) and (\ref{Delt1})
we have used 12-electron version of the GRECP operator for Hf used earlier in Refs.\cite{Petrov:07a,Petrov:09b,Skripnikov:08a} to perform 2-component 20-electron correlation calculations. For the calculations we have used the [12,16,16,10,8]/(6,5,5,3,1) basis set for Hf and [14,9,4,3]/(4,3,2,1) ANO-I basis set for F \cite{Roos:05}. Calculations of the matrix elements 
%(\ref{Gperp1}), (\ref{Delt1}) and (\ref{dip1}) 
(\ref{Gperp1}), (\ref{dip1}) and (\ref{Delt1})
were performed within the linear-response coupled cluster with single and double cluster amplitudes, LR-CCSD, method.
%ls
%The matrix elements (\ref{Gperp2}), (\ref{Delt2}) and (\ref{dip2}) were calculated using the \textit{multireference} linear-response coupled cluster with single and double cluster amplitudes, LR-MRCCSD, method. Active spinor space included leading $\sigma$ and $\pi$ orbitals of HfF$^+$ which are involved in the $^3\Delta_1 -- ^3\Pi_{0^\pm}$ transition.
%end ls

%ls1
%Hartree-Fock calculations were performed using the {\sc dirac12} code \cite{DIRAC12}. Relativistic correlation calculations were performed within the {\sc mrcc} code  \cite{MRCC2013}. 
%For scalar-relativistic calculations we used the {\sc cfour} code \cite{CFOUR,Gauss:91,Gauss:93,Stanton:97}.
Electronic calculations were performed within the \cite{DIRAC12} and \cite{MRCC2013} codes.
%end ls1
The code to calculate matrix elements of the g-factor operator over the 4-component molecular bispinors has been developed in the present paper.

\section{Results and discussions}
%aptav
$G_{\parallel}$ obtained from the electronic structure calculation is equal to 0.0115 and is in avery good agreement with the value $G_{\parallel}=0.011768$ obtained by fitting the $g_{\rm fit}=-0.00306$ value. In Ref. \cite{Cornell:13} the experimental value $g_{\rm exp}=+0.00306$
obtained in the external electric field $\cal{E}=$11.6 V/cm is given.
The electronic structure calculation is in agreement with the experiment only if
the sign of $g-$factor will be changed.
Thus, for consistency with the experiment, in this work we furhter use the g-factor value
$g_{\rm fit}= -g_{\rm exp}$  with the sign reversed from that in Ref.  \cite{Cornell:13}.
%end aptav
Only $G_{\parallel}$ parameter was optimized in the fitting procedure.
Eq. (\ref{geq}) gives $G_{\parallel}$ = 0.012043.
%Thus, nonadiabatic corrections are rather 

\begin{figure}
\includegraphics[width = 3.3 in]{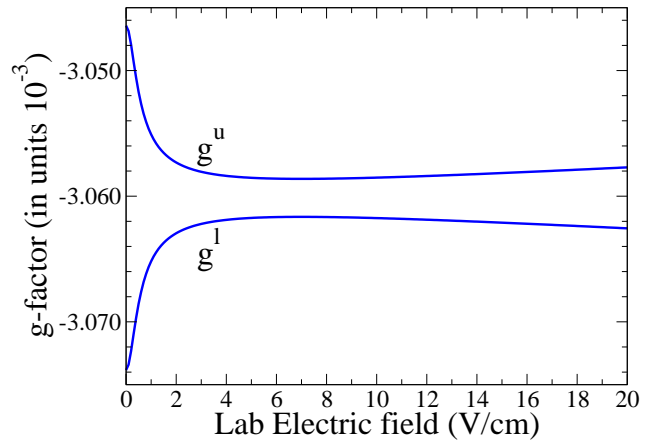}
 \caption{(Color online) Calculated $g-$factors for upper ($g^u$) and lower ($g^l$) levels of $\Omega$ doublets for the $J = 1, F=3/2, M_F=3/2$ levels of the $^3\Delta_1$ state of HfF$^+$ as functions of the electric field.}
 \label{gfgecross}
\end{figure}

In Fig.~\ref{gfgecross} the calculated $g$-factors for the $J = 1, F=3/2, M_F=3/2$ levels of HfF$^+$ $^3\Delta_1$ state are shown as functions of the laboratory electric field.
The calculated difference $\Delta g = g^u - g^l = 3.4\times10^{-6}$ between the g-factors of the upper ($g^u$) and lower ($g^l$) levels of $\Omega$ doublets
 is in a good agreement with the experimental value $-1(2)\times10^{-5}$ \cite{Cornell:13}.
Note, that the difference is zero in the adiabatic approximation. 
%ap
The lowest
value for the difference, 
$\Delta g = 3 \times 10^{−6}$, is attained at the electric field \cal{E}=7 V/cm.
The smaller $\Delta g$, the smaller are
the systematics $\sim \mu_B\Delta g\tilde{B}$ coming from a spurious magnetic 
field $\tilde{B}$.
%end ap

%\begin{figure}[pH]

%\section{Conclusion}

\section{Acknowledgement}
Molecular calculations were partly performed on the Supercomputer ``Lomonosov''.
The development of the code for the computation of the matrix elements of the considered operators as well as  the performance of all-electron calculations were funded by RFBR, according to the research project No.~16-32-60013 mol\_a\_dk. GRECP calculations were performed with the support of President of the Russian Federation Grant No.~MK-7631.2016.2 and Dmitry Zimin ``Dynasty'' Foundation.
The calculations of $g-$factor dependence on electric field  were supported by the grant of Russian Science Foundation No.~14-31-00022.

%\bibliographystyle{./bib/apsrev}
%\bibliography{bib/JournAbbr,bib/SkripnikovLib,bib/QCPNPI,bib/TitovLib,bib/Kaldor,bib/PetrovLib,bib/Lomachuk,bib/Kudashov,bib/ACME}

\end{document}